\documentclass[useAMS,usenatbib]{mn2e}
\usepackage{amsmath}
\usepackage{amsbsy}
\usepackage{times}
\usepackage{graphicx}
\usepackage{aas_symbols}
\usepackage{epstopdf}

\usepackage{cases}
\usepackage{color}

\newcommand{\beq}{\begin{equation}}
\newcommand{\eeq}{\end{equation}}
\newcommand{\bea}{\begin{eqnarray}}
\newcommand{\eea}{\end{eqnarray}}
\newcommand{\gae}{\lower 2pt \hbox{$\, \buildrel {\scriptstyle >}\over {\scriptstyle
\sim}\,$}} 
\newcommand{\lae}{\lower 2pt \hbox{$\, \buildrel {\scriptstyle <}\over {\scriptstyle
\sim}\,$}}

\usepackage{ulem}

\begin{document}

\title[Off-axis GRBs from GW events]{Off-axis short GRBs from structured jets as counterparts to GW events}

\author[Kathirgamaraju, Barniol Duran \& Giannios]{Adithan Kathirgamaraju$^{1}$\thanks{Email: akathirg@purdue.edu (AK), barniolduran@csus.edu (RBD), dgiannio@purdue.edu (DG)}, Rodolfo Barniol Duran$^{2}$\footnotemark[1], Dimitrios Giannios$^{1}$\footnotemark[1] \\
$^{1}$Department of Physics and Astronomy, Purdue University, 525 Northwestern Avenue, West Lafayette, IN 47907, USA\\
$^{2}$ Department of Physics and Astronomy, California State University, Sacramento, 6000 J Street, Sacramento,  CA 95819, USA 
}

\date{Accepted to MNRAS Letters}

\pubyear{2017}

\maketitle

\begin{abstract}
Binary neutron star mergers are considered to be the most favorable
sources that produce electromagnetic (EM) signals associated with
gravitational waves (GWs). These mergers are the likely progenitors of
short duration gamma-ray bursts (GRBs). The brief gamma-ray emission
(the ``prompt" GRB emission) is produced by ultra-relativistic jets, as a
result, this emission is strongly beamed over a small solid angle
along the jet. It is estimated to be a decade or more
before a short GRB jet within the LIGO volume points along our line of
sight. For this reason, the study of the prompt signal as an EM
counterpart to GW events has been sparse. We argue that for a
realistic jet model, one whose luminosity and Lorentz factor vary smoothly
with angle, the prompt signal can be detected for a
significantly broader range of viewing angles. This can lead to an ``off-axis" short GRB as an EM counterpart. Our estimates and
simulations show that it is feasible to detect these signals with the
aid of the temporal coincidence from a LIGO trigger, even if the
observer is substantially misaligned with respect to the jet. 
\end{abstract}


\section{Introduction}
The monumental discovery of gravitational waves by the LIGO collaboration enables us to observe our Universe at a new wavelength (\citealp{abbott2016c,abbott2016b}). In particular, 
gravitational waves allow us to study the merger of compact objects and their properties, offering 
exquisite tests of general relativity (\citealp{abbott2016a}). The next related major quest in astronomy 
is the discovery of an electromagnetic signal produced during such a merger that accompanies 
the gravitational waves. We will refer to this specific electromagnetic signal as an electromagnetic
``counterpart." 

By far the most promising source of gravitational waves (GWs) and accompanying electromagnetic 
(EM) signals are double neutron star (NS-NS) mergers (or neutron star-black hole mergers), hereafter 
referred to as simply ``mergers" (e.g., \citealp{lee2007}).  Such mergers make for promising detectable GW sources by LIGO 
within a few hundreds of Mpc (\citealp{martynov2016}). There are several lines of indirect evidence that 
suggests these mergers are the most likely progenitors of short GRBs 
(\citealp{eichler1989, nakar2007, berger2014}). However, a simultaneous GW and GRB detection would 
provide a most conclusive evidence that short GRBs are indeed produced during binary mergers.  

The ``prompt" $\gamma$-ray emission from short GRBs is believed to be strongly beamed along an ultra-relativistic jet with half opening angle $\theta_{\rm j}$ and Lorentz factor $\Gamma_{\rm core} \gae 30$ (e.g., \citealp{nakar2007}). If $\Gamma_{\rm core}\theta_j>1$, it will be extremely difficult to detect the prompt emission from a short GRB jet that is misaligned by an angle $\theta>\theta_{\rm j}$ with respect to Earth. 
In fact, the observed rate of short GRBs indicates that it will be a decade or more before 
the luminous core of a GRB is detected within the LIGO detectability volume of neutron star 
mergers (e.g., \citealp{wandermanandpiran2015}). This has tended to steer investigations of 
EM counterparts away from the prompt emission (see however \citealp{kochanek1993,patricelli2016,Lazzati2017}), and more towards the less 
prompt signals that follow days to months after the merger/GW detection; 
such as ``macronova" or ``kilonova", off-axis afterglows and radio flares 
(e.g., \citealp{li1998, metzger2010, metzgerandberger2012, kasen2013, nakarandpiran2011, 
hotokezakaandpiran2015,lamb2017}). Given the poor localization of LIGO (\citealp{abbott2016d}), the 
faintness of these signals, and their long delay, such detections and their association to the merger 
will be challenging (e.g., \citealp{metzgerandberger2012}).

Here we investigate a different, prompt signal from the merger; that
of the prompt emission from the moderately relativistic $\Gamma\sim$
a few part of the jet, the ``sheath'', that beams its emission towards the observer (who is located at a 
substantial angle with respect to the jet's core). In this letter, we argue that by exploiting 
the timing from a LIGO trigger, one can reliably detect the prompt emission even 
if the jet is significantly misaligned with respect to Earth. This is because, in any realistic jet 
model, there is expected to be a slower, under-luminous 
sheath surrounding the bright jet core (e.g., \citealp{rossi2002,salafia2015}). To quantify this claim, we perform large-scale relativistic 
magnetohydrodynamical (MHD) simulations, which follow the jet from the launching region, 
through the confining ambient gas and the break out distance where its slower sheath forms. We also provide a calculation  
to estimate the observed luminosity for an observer located at an arbitrary angle with respect to the 
jet axis. We find that this endeavor is quite promising.

\section{Our Model: a structured jet}

Just prior to the merger of a binary neutron star system, gravitational and hydrodynamical interactions expel 
some neutron star material, forming the ``dynamical ejecta"  (e.g.,
\citealp{hotokezaka2013, rosswog2013}). The neutron star merger may be
followed by the launching of an ultra-relativistic jet.  As investigated by previous hydro simulations, the jet is initially collimated by the dynamical ejecta
until it breaks out from the surrounding gas (\citealp{nagakura2014,berthier2014,duffell2015}).  At a larger distance, it dissipates its energy, 
resulting in a short GRB which lasts for $\lesssim 2$ s and peaks at $\sim$ MeV energies 
(\citealp{nakar2007, berger2014}). In the majority of previous models,
this jet consists of a core having uniform luminosity ($L_{\rm core}$) and Lorentz factor ($\Gamma_{\rm core}$) 
that discontinuously disappears for angles $\theta>\theta_{\rm j}$. However, these models are not physical and 
greatly underestimate the prompt emission that may be received by observers who are not 
aligned within the core of the jet (i.e., off-axis observers).

\begin{figure}
 \begin{center}
 \includegraphics[height=6.0cm,angle=0]{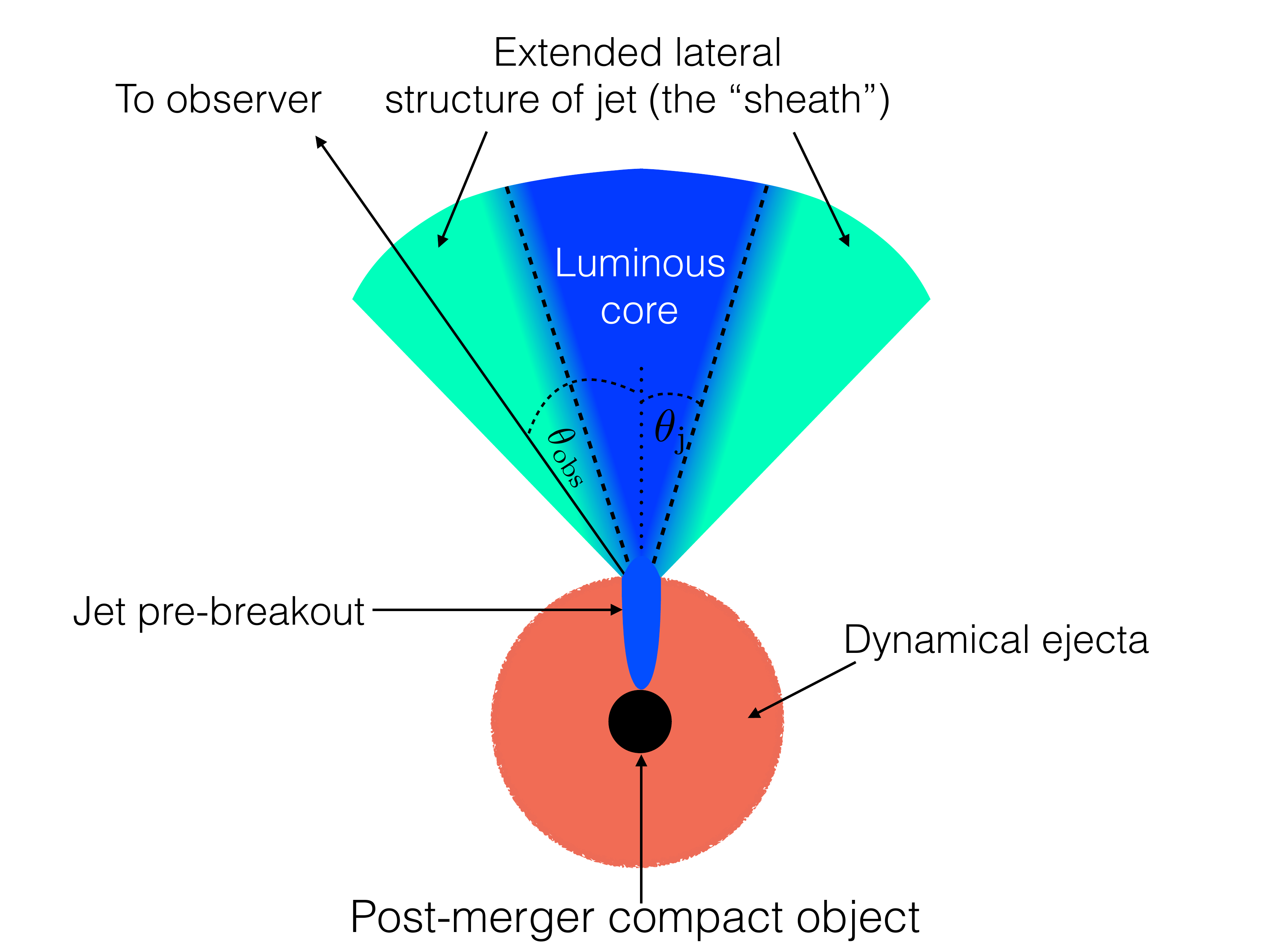}
 \end{center}
\caption{A schematic of a short GRB jet. Mergers produce 
GWs detectable by LIGO and are the likely progenitors of short GRBs. The prompt emission 
from the jet's luminous core (routinely observed as a short GRB) is strongly 
beamed and can only be detected by observers located within $\theta_{\rm j}$ from the jet 
axis. However, the jet is expected to have a lateral structure that moves slower and is 
fainter than the luminous core. Given the proximity of a LIGO-triggered short GRB, {\it Fermi} 
and {\it Swift} can potentially detect the prompt emission from this lateral structure even if the jet is 
misaligned with respect to our line of sight (see Section \ref{feasibility}).}
\label{sgrbjet}
\end{figure}

Recent numerical (e.g., \citealp{sasha2010,komiss2010}) and theoretical 
(e.g., \citealp{sapountzis2014}) studies show that once a magnetized jet breaks out of the collimating medium, it is 
expected to develop some ``lateral structure". This means that the jet's luminosity and Lorentz factor depend on the polar angle $\theta$ (see Fig. \ref{sgrbjet}). We find that this extended lateral part, though slower and less luminous, can contribute a significant amount of prompt emission for angles larger than $\theta_{\rm j}$. As a result, it is possible to
detect the prompt emission from a structured jet for observing angles ($\theta_{\rm obs}$) much larger than $\theta_{\rm j}$. We call this prompt emission detected by observers with $\theta_{\rm obs}>\theta_{\rm j}$ an ``off-axis" short GRB, this is not off-axis as defined in the traditional sense because our jet does not abruptly vanish at $\theta_{\rm j}$. Below we provide an estimate for
the prospects and feasibility of detecting this ``off-axis"
prompt emission and mention some of its advantages over other EM counterparts.

\subsection{Feasibility of detecting the prompt emission from a structured jet} \label{feasibility}

There are currently more than thirty short GRBs with measured redshift (\citealp{fong2015}), 
and their average redshift is $\sim$0.5 (\citealp{berger2014}). Let us now pick a typical short GRB with 
known redshift, assume it takes place within the LIGO detectability volume, and estimate its off-axis
prompt emission. Had short GRB101219A (see table \ref{table1}) taken place at a distance of $\sim$ 200 Mpc, it 
would have resulted in an extremely bright source with a count rate of $\sim10^6$ photons/s 
at the {\it Fermi}/GBM detector. This is a factor of $f\sim$$10^4$ above the count rate required 
for a robust detection of a source coincident with a LIGO trigger by {\it Fermi} 
(\citealp{connaughton2016}). With such a large on-axis count rate, even a steeply 
declining luminosity for the lateral structure of the jet will provide a significant amount of 
off-axis emission that can be detectable by, e.g., {\it Fermi}. Assuming, for
the sake of an estimate, a jet with a core of luminosity 
$L_{\rm core}$ and half opening angle $\theta_{\rm j}$, we can take the luminosity for angles 
$\theta>\theta_{\rm j}$ (i.e for the lateral structure of the jet) to drop sharply as 
$L_{\rm obs}(\theta)=L_{\rm core} (\theta/\theta_j)^{-6}$ (\citealp{pescalli2015}). 
Such a jet can still be detected by an observer up to an angle
$\theta_{\rm obs} \sim f^{1/6}\theta_{\rm j}\sim 5 \theta_{\rm j}$. 
This makes it $(\theta_{\rm obs}/\theta_{\rm j})^2 \sim 20$ times more
likely to observe the sheath of the jet in comparison to its core
emission. Instead of detecting about 1 EM counterpart per decade from
the prompt core emission,  the sheath potentially results in $\sim$a
few events per year for an instrument with {\it Fermi's} field of view, increasing the 
chances to detect such events tremendously. Therefore, the prompt emission from a short 
GRB could be detected even if the jet is significantly misaligned.

It is quite likely that the off-axis $\gamma$-ray emission, by itself, is not sufficiently bright enough to result in 
a detector trigger. Nevertheless, using the timing of a LIGO trigger can make even a faint 
$\gamma$-ray signal a statistically significant detection. A faint $\gamma$-ray signal must come within 
several seconds after a LIGO trigger to make such a detection possible
(\citealp{connaughton2016}). Here, we estimate the temporal difference of these signals. 
The GW signal, as detected by LIGO, is expected to peak approximately when the merger takes place.  The merger will probably initially give birth to a fast-rotating, massive proto-neutron star 
and can take $\sim$ hundred milliseconds to collapse to a black hole (e.g., \citealp{rezzolla2011}). It is 
likely that the jet forms a few dynamical times later or, all in all, $\sim 0.1-1$~sec after the
GW signal peak, then the jet has to breakout of the collimating medium which can take a few hundreds of msec (\citealp{nagakura2014}), and expand to a radius $r_{\rm jet}$, where it radiates. The emission 
from the jet will, therefore, be further delayed by $r_{\rm jet}/\Gamma^2c$, where and $\Gamma$ is the Lorentz factor of the patch of the jet directed towards the observer. The fast rise and 
variability of short GRBs indicates the jet core is characterized by
$r_{\rm jet}/\Gamma_{\rm core}^2c\sim10$ msec. In Section \ref{discussion}, we argue
that the sheath emission is also likely characterized by a delay of
$r_{\rm jet}/\Gamma^2c\sim$a few seconds, i.e., making a very prompt signal.

The misaligned or ``off-axis" prompt emission of short GRBs has largely been ignored. The community has 
rather focused its efforts on studying off-axis afterglows from short GRBs, the ``macronova/kilonova" and 
radio flare produced by the dynamical ejecta, and the emission from other components 
(e.g., the cocoon emission; \citealp{nakarandpiran2017, gottlieb2017}). These signals 
are expected to peak $\sim$ days to $\sim$ years after the merger and
are fainter compared to the prompt emission
 (e.g., \citealp{metzgerandberger2012, metzger2017}). Coupled with the fact that current GW detectors have very 
poor localization, associating these signals to a GW source will be difficult. This is where the ``off-axis" prompt 
signal has a considerable advantage, since we expect to detect the prompt emission a few seconds after the 
GWs, we can capitalize on its temporal coincidence to make the detection. This will also make follow-up 
searches for the off-axis afterglow and macronova/kilonova easier. Hence, it is very likely that the first detected 
EM counterpart of a LIGO trigger involving a NS-NS merger will be the fainter, ``off-axis" prompt emission from a 
short GRB jet.

\begin{table}
\begin{center}
\begin{tabular}{ccccc}
\hline
GRB & $T_{90}$ & Fluence & $\beta$ & $z$\\
\hline
101219A & 0.6 s & $4.6\times10^{-7}$ erg cm$^{-2}$ & 0.6 & 0.72\\
\hline
\end{tabular}
\end{center}
\caption{Observed parameters of GRB 101219A. From left to right, the columns indicate: GRB identifier, burst duration, fluence in the 
15-150 keV band, photon-number index $\beta$ as a function of frequency 
$\rm {d}N/\rm{d}\nu \propto \nu^{-\beta}$, and the burst's redshift.  In Section \ref{feasibility}, 
we use this particular GRB to show that a typical short GRB placed at 200 Mpc (LIGO volume) 
will be very bright. Consequently, the ``off-axis" prompt emission could be detectable even for substantially misaligned observers. This increases the likelihood to detect these objects, 
using a LIGO trigger, even in the absence of a GRB trigger (data from \citealp{ryan2011, fong2015}).}
\label{table1}
\end{table}


\section{Numerical simulations} \label{mhd_simulations}

We have recently run relativistic MHD simulations of AGN jets (\citealp{barniolduran2017}; hereafter BTG17) 
using the HARM code (\citealp{gammie2003}), with recent improvements (\citealp{sasha2011, 
omerandsasha2016}). 
The initial conditions and numerical scheme of these simulations are adapted to the physical setup 
relevant to this work. We initiate the jets via the rotation of the central magnetized compact 
object. The jets are, therefore, launched magnetically dominated. By adjusting the density of the gas in the injection 
radius, the jet is launched with magnetization $\mu = 2p_{\rm mag,0}/\rho_0 c^2 \simeq 25$, where $p_{\rm 
mag,0}$ is the magnetic pressure and $\rho_0$ is the density at the base of the jet.  The initial
magnetization $\mu$ determines maximum Lorentz factor of the jet. We do not take into account neutrino heating which may affect the structure of the jet (e.g., \citealp{barkov2011,berthier2017})

 We considered a scenario in which 
a low-density funnel is carved along the z-axis at the start of the simulation and we have 
confined the jet to propagate along high-density walls (similar to ``model B" simulations in BTG17). 
Our simulations, performed in both 2D and 3D, follow jets from the compact object to a scale $\sim 
10^3 \times$ larger. These large scale simulations allow us to follow the jet acceleration through conversion 
of its magnetic energy into kinetic energy. 
In the context of short GRB jets, as the jet breaks out from the confining medium, it essentially 
travels through vacuum (or at least very low ambient gas). This is quite advantageous since 
BTG17 have shown that these type of jets are almost identical in 2D and 3D runs 
and that they are fairly axisymmetric (see also \citealp{omerandsasha2016}). For this reason, we focus our 
efforts on axisymmetric 2D simulations, which are considerably less computationally 
expensive and can be better resolved numerically. 

\begin{figure}
\begin{center}
\includegraphics[height=6.75cm,angle=0]{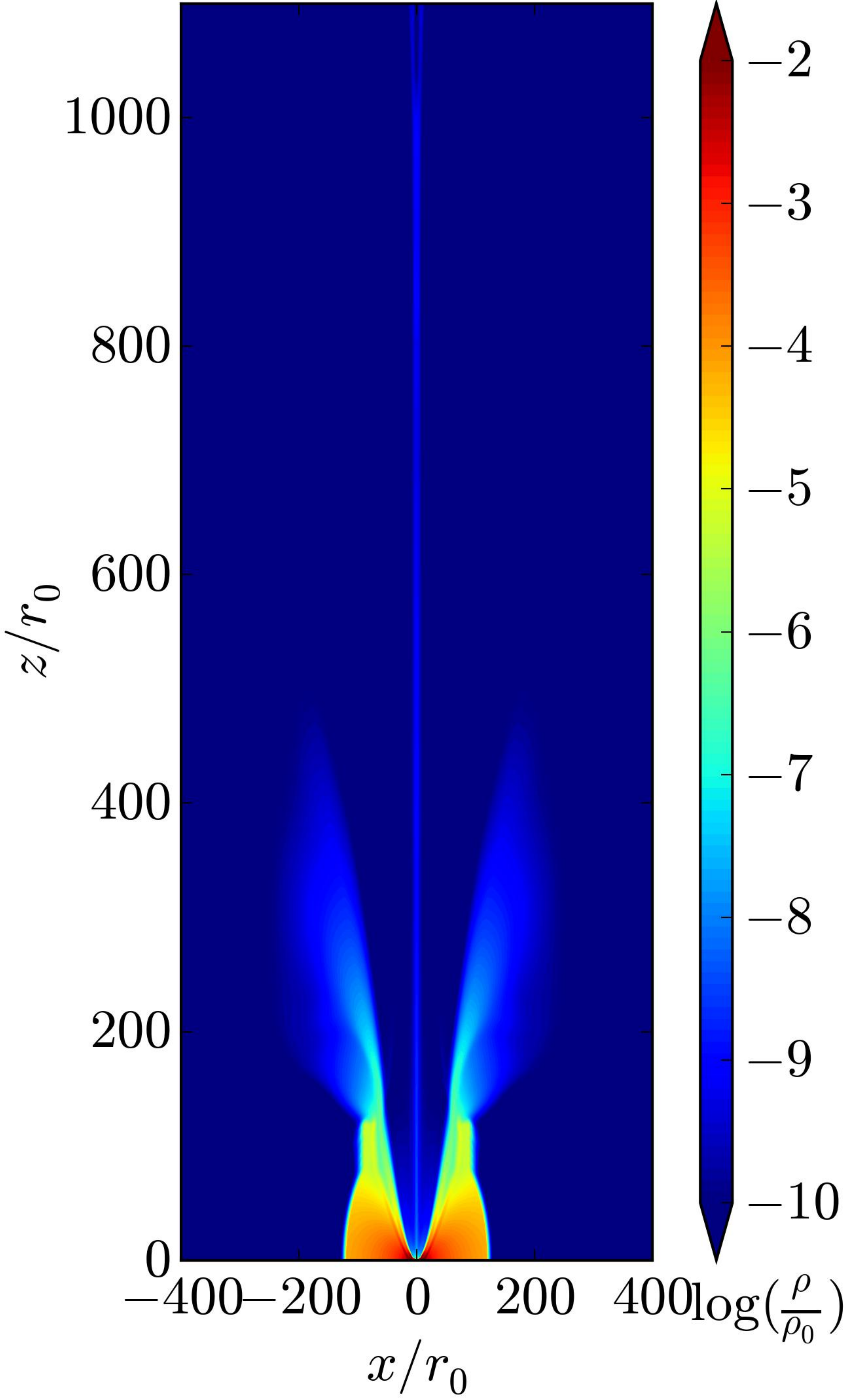}
\includegraphics[height=6.75cm,angle=0]{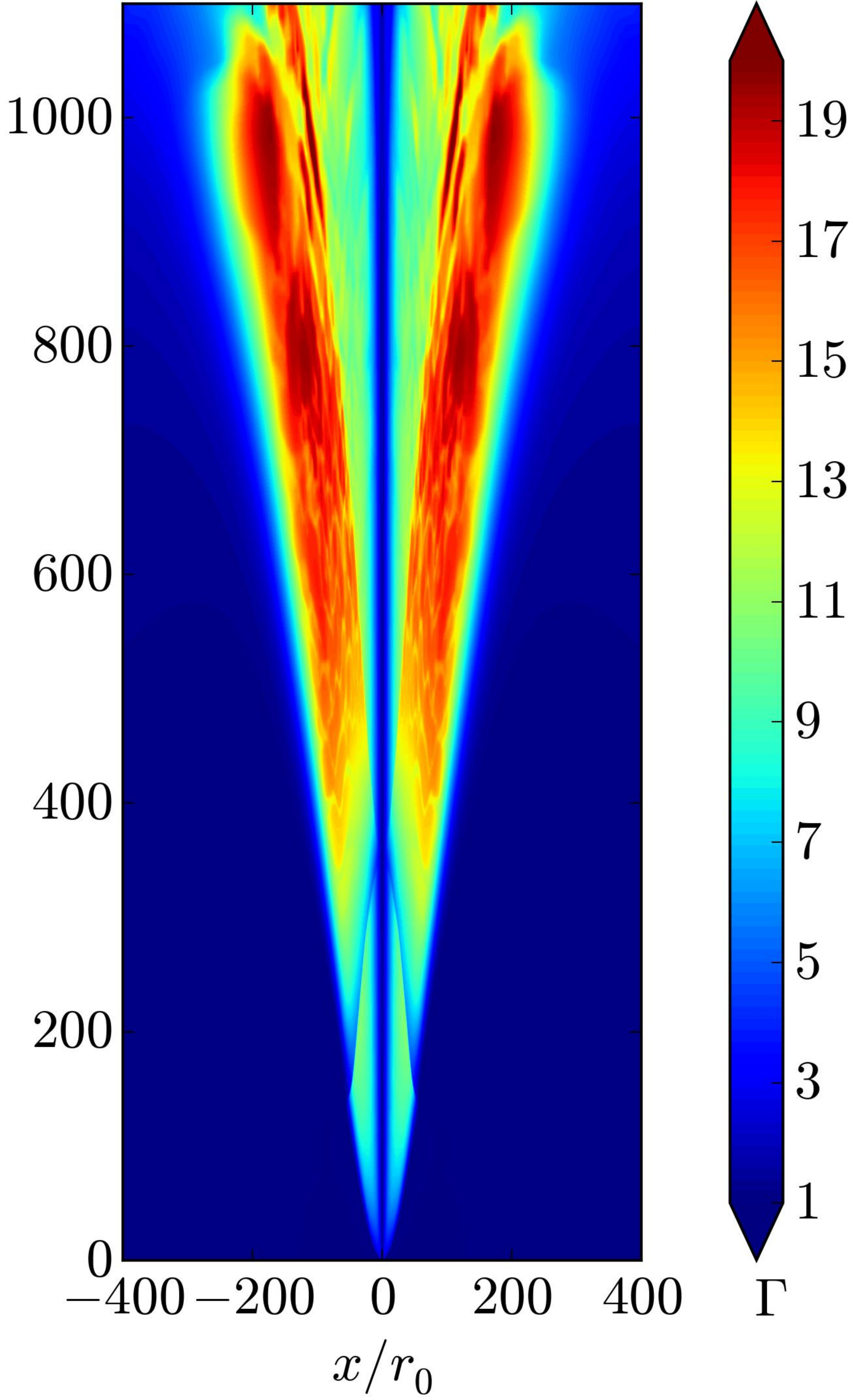}  \\
\end{center}
\caption{Numerical simulation of a jet that is collimated by and 
breaks out from the dynamical ejecta. We 
show 2D cuts of density (left panel) and Lorentz factor (right panel),
where $r_0$ stands for a few times the radius of the central compact
object. The jet accelerates
as it breaks out from the dynamical ejecta and spread sideways. At
large distance the jet turns conical and its lateral structure is
fixed.}
\label{figure_test}
\end{figure}

We mimicked the boundary of the dynamical ejecta by setting the ambient gas density
of our previous simulations (see BTG17 model B-2D-vhr for details on jet and ambient gas 
parameters and numerical resolution) 
to be essentially zero beyond $\sim 100$ times the size of the compact 
object as displayed in Fig. \ref{figure_test}, which shows both density and velocity (Lorentz factor) contours. 
After the jet breaks out from the dynamical ejecta, a rarefaction wave 
crosses the jet and it spreads sideways and accelerates further (\citealp{sasha2010,komiss2010}).  
For this work, we simulate ``steady state" jets that will maintain the conditions at the central engine 
constant. This means that the rotation of the black hole and the magnetic field strength
is kept constant, so that the jet has a constant power throughout
the duration of the simulation. Future simulations will explore short-lived jets.

The distance where the radiation from the jet takes place is
uncertain; the prompt emission may originate at the transparency 
radius of the jets or further out, at optically thin conditions (see
next Section). The distance at which the  jet produces the $\gamma$-ray
radiation lies beyond our simulated region. However, we argue that
the dynamical range of our current simulations is sufficient for the
objectives of our estimates.
This is due to the fact that after the phase of lateral expansion and 
the crossing of the rarefaction wave, the jet becomes approximately 
conical and its properties ``freeze out" as a function of angle (see Fig. \ref{figure_LandGamma}). Therefore, the
lateral structure of the jet, which is the key aspect of our study,
is set at this distance. It is no longer necessary to follow 
the jet beyond this point since its properties can be safely
extrapolated at larger distances (\citealp{sasha2010}). Our current simulations 
with a dynamical range of $\sim 10^3$ already point towards this jet feature
after breakout. Future simulations in 2D with even larger dynamical range will 
be able to test this effect in detail.

\section{Results and Discussion}
\label{discussion}

\begin{figure}
\includegraphics[height=6.0cm,angle=0]{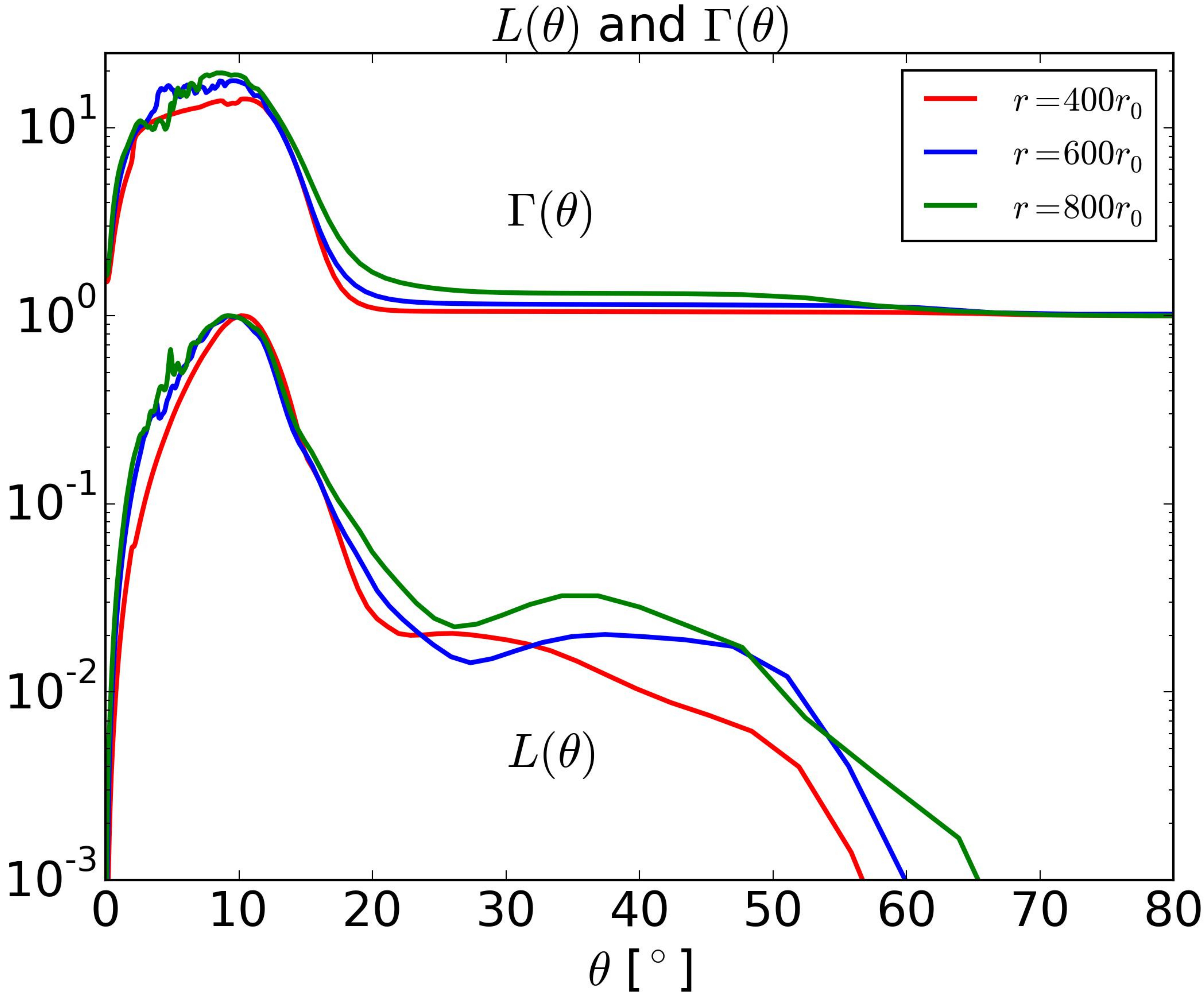} 
\caption{Jet luminosity, $L(\theta)$ (arbitrary units), and jet Lorentz factor, 
$\Gamma(\theta)$, for different radii $r$ from the compact object for our 
numerical simulation in Fig. \ref{figure_test},  
$r_0$ stands for a few times the radius of the central compact
object. The
luminosity and Lorentz factor profiles are very similar for increasing
radii, hence we can assume that the jet structure "freezes out" beyond
a certain radius. 
This allows us to safely extrapolate the jet structure to even larger radii.}
\label{figure_LandGamma}
\end{figure}

We have extracted the jet luminosity $L(\theta)$ and Lorentz factor $\Gamma(\theta)$  from our simulation
in Fig. \ref{figure_test} at different radii from the
central object, see Fig. \ref{figure_LandGamma}.
The luminosity $L(\theta)$ is the total (magnetic, kinetic
and thermal) luminosity per solid angle of the jet
$L(\theta)=dL/d\Omega$.  We note that close to
the jet axis (z-axis) the luminosity is very low and the velocity is
quite small. The jet is characterized by a fast and luminous core of
opening angle of $\theta_{\rm j}\sim 10^{\rm o}$. The typical
cosmological GRB is observed through its core emission. However, for
larger angles, the jet Lorentz factor and luminosity drop steeply but remain substantial. These features have been seen in MHD
jets before (see, e.g., \citealp{sasha2008} and references therein), but
the exact profiles of  
$L(\theta)$ and $\Gamma(\theta)$ should depend on their profiles right
before breakout, which in turn depend on the properties of the
dynamical ejecta (see Section \ref{mhd_simulations}). Future simulations 
will explore more realistic dynamical ejecta models (e.g., \citealp{hotokezaka2013, nagakura2014}). 
Recent hydrodynamical simulations also find an energetic component at 
large angles (e.g., \citealp{gottlieb2017}), and this component is ascribed 
to the ``inner" cocoon (shocked jet material). However in our simulations, the effect of 
the cocoon on the jet structure is minimized because we initiate our setup with an evacuated funnel, and the extended component (the ``sheath") we obtain consists
of rarefied magnetized jet.

As seen in Fig. \ref{figure_LandGamma}, far beyond the breakout radius, there is evidence that
quantities ``freeze out",
i.e., the jet shows a similar profile as a function of angle for increasing radii. We will make use of this fact and
show how the observed luminosity can be extracted from our MHD
simulations. 

\subsection{Calculating observed luminosity}

\begin{figure}
\includegraphics[height=5.5cm,angle=0]{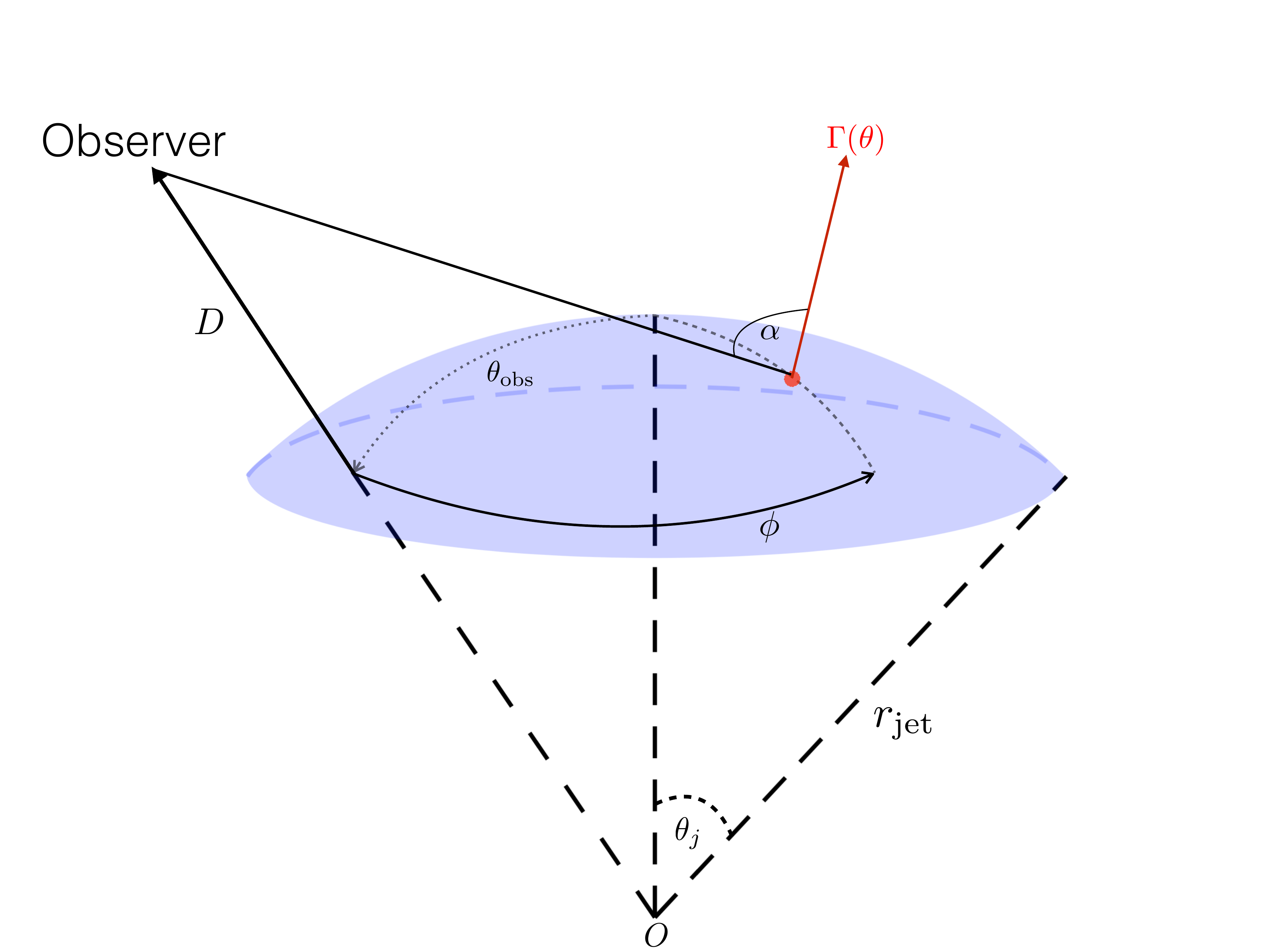}\\
\includegraphics[height=5.0cm,angle=0]{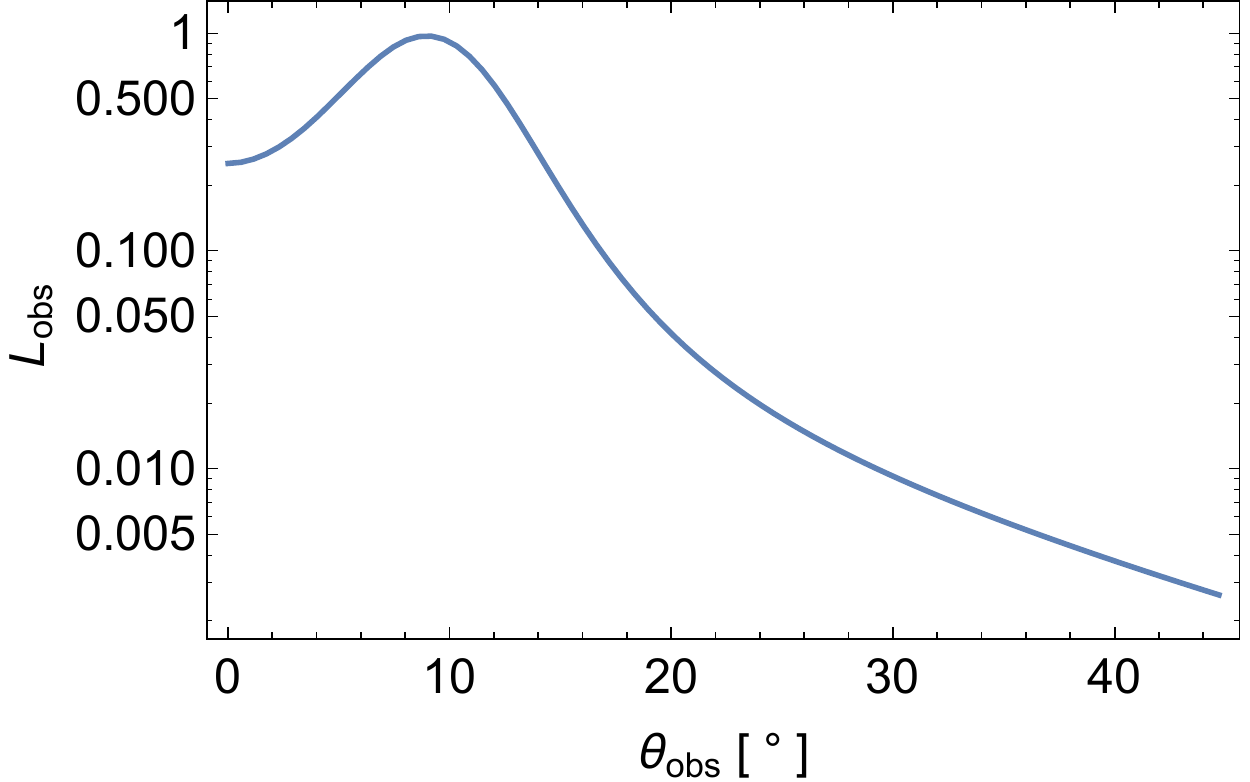}
\caption{{\it Top panel:} Geometry of the conical jet, we use spherical coordinates with the 
origin at $O$. The observer is located at ($D$, $\theta_{\rm obs}$, 0). We consider a patch of this jet 
(red dot) at ($r_{\rm jet}$, $\theta$, $\varphi$) moving with Lorentz factor $\Gamma(\theta)$ with a 
corresponding angle $\alpha$ between its velocity and line of sight of
the observer.  {\it Bottom panel:} Observed luminosity (normalized to peak) as a function of 
observer angle, $L_{\rm obs}(\theta_{\rm obs})$, for our jet simulation output presented in 
Fig. \ref{figure_LandGamma}. The calculation was performed at $r=800
r_0$, $r_0$ is a few times the size of the compact object.  In this example, the jet luminosity at $\sim 40^{\rm o}$ is a
factor of 300 fainter than that of the jet core. Nevertheless, such a misaligned jet
can be detected by a $\gamma$-ray instrument if it takes place within the Advanced LIGO detectability volume.}
\label{jetschematic}
\end{figure}

We now briefly describe how to calculate the observed luminosity as a function of $\theta_{\rm obs}$ for a given 
$L(\theta)$ and $\Gamma(\theta)$. All quantities in the local co-moving frame 
will be denoted with a prime and we employ spherical coordinates. Consider an infinitesimal patch of the jet located at a polar angle $\theta$ from the jet axis and azimuthal angle of $\varphi$ (where we take the observer to be located at $\theta_{\rm obs}$ and $\varphi=0$, see Fig. \ref{jetschematic}). The portion of the jet within this patch moves with 
$\Gamma(\theta)$ at an angle $\alpha$ with respect to the line of sight of the
observer, where $\alpha$ is given by
${\rm cos\, \alpha}={\rm cos\, \theta_{\rm obs}\, cos\, \theta+sin\, \theta_{\rm obs}\, sin\, \theta \, cos\, \varphi}$.

Suppose this patch subtends a solid angle 
$d\Omega_{\rm p}$ on the jet-surface, the luminosity through this patch will be 
$L_{\rm p}=L(\theta)d\Omega_{\rm p}$. Using a Lorentz transformation, the luminosity within 
this patch in the co-moving frame can be expressed as $L_{\rm p}'= L_{\rm p}/\Gamma^2(\theta)$. We assume a fixed fraction ($\eta$) of this luminosity is
converted into radiation. For simplicity, we further assume that the 
radiation is released instantaneously, is isotropic in the jet co-moving frame and is emitted at a 
fixed distance $r_{\rm jet}$, which is justifiable since the jet structure ``freezes out" at the distances we are considering, therefore the total luminosity from a gradual dissipation would give similar results. The radiated 
luminosity per unit solid angle in the co-moving frame is therefore $\eta L_{\rm p}'/4\pi$. This luminosity per solid angle has to 
be boosted  to the lab frame, taking into account the inclination $\alpha$. Therefore, each patch of the jet 
contributes
\beq
dL_{\rm obs}= \Gamma(\theta)\delta^3\frac{\eta L_{\rm p}'}{4\pi}=\frac{\eta L(\theta)d\Omega_{\rm p}}{4\pi\Gamma^4(\theta)[1-\beta(\theta) {\rm cos}\,\alpha]^3}.
\label{dlobs}
\eeq
Finally, we add the 
contribution from all patches of the jet which amounts to an integral over the solid angle of the jet, hence,
\beq
L_{\rm obs}(\theta_{\rm obs})= \int_0^{2\pi}\int_0^{\theta_{\rm e}}  dL_{\rm obs},
\label{lobs}
\eeq
where $\theta_{\rm e}$ signifies the poloidal extent of the jet. This calculation shows that by extracting $L(\theta)$ and $\Gamma(\theta)$ from our simulations we can 
estimate the prompt emission seen by an observer at any angle. The observed luminosity of the core can be scaled to the 
observed average luminosity of on-axis short GRBs, which in turn can give us a count rate at a detector (e.g., {\it Fermi} or {\it Swift}).
Fig. \ref{jetschematic} shows the observed luminosity $L_{\rm obs}(\theta_{\rm obs})$ (normalized to the peak luminosity)  for our simulation 
in Fig. \ref{figure_test}, using $L(\theta)$ and $\Gamma(\theta)$ shown in Fig. \ref{figure_LandGamma}, taking $\theta_{\rm e}\sim 23^{\circ}$, which marks the extent of the jet with substantial magnetization ($\sigma\gtrsim10^{-2}$). 
The observed luminosity decreases quickly as the angle from the jet
axis increases; however, at large angles a significant contribution
exists. In this example, the jet luminosity at $\sim 40^{\rm o}$ is a
factor of $\sim$300 fainter than that of the jet core. Nevertheless, it
is clear from the estimates presented in Section 2.1 that such a
misaligned jet can be still detected, provided that it takes place within the Advanced LIGO detectability volume.

If we had considered the uniform core model for the jet with
$\Gamma_{\rm core} \approx 20$, the ratio of the observed luminosity at
$\theta_{\rm obs} \approx 40^{\rm o}$ to the on-axis ($\theta\lesssim
10^{\rm o}$) luminosity would be $L(\theta_{\rm obs})/L_{\rm core}\simeq (\Gamma_{\rm core}\theta_{\rm obs})^{-6}\sim
10^{-7}$, which would be negligible. Hence, the ``off-axis" prompt emission 
from a structured jet is significantly larger than that from the uniform jet model, which greatly increases
the prospects of detecting it.

The {\it steady jet} assumption considered above is valid as long as the GRB duration 
(defined for an on-axis observer) is $T_{\rm GRB} > r_{\rm jet}/\Gamma^2 c$, where 
$r_{\rm jet}$ is the radius at which the jet dissipation occurs, and the 
$\gamma$-rays for an on-axis observer are produced. If this condition is not satisfied 
then (i) the onset of the emission is delayed by $\sim r_{\rm jet}/\Gamma^2 c$ and 
(ii) the luminosity drops by a factor of $\sim\Gamma^2 c T_{\rm GRB}/r_{\rm jet}$ 
with respect to the steady jet calculation performed above. This evidently depends on 
$r_{\rm jet}$ and $\Gamma(\theta)$, and therefore on a particular jet
dissipation model. For an estimate, we will consider the 
 {\it photospheric} model (e.g., \citealp{meszaros2000, giannios2006})
for the prompt GRB emission. This
model predicts that the emission comes from the Thomson photosphere of
the jet: $r_{\rm jet}=r_{\rm ph}=L \sigma_T/4\pi \Gamma^2 \mu m_p
c^3$. The corresponding delay of the prompt signal will be $\sim
r_{\rm jet}/2\Gamma^2 c \simeq L \sigma_T/8\pi \Gamma^4 \mu m_p
c^4\sim 5 L_{48}\Gamma_{0.5}^{-4}\mu_{1.5}^{-1}$~sec, where we use
the notation $A=10^xA_x$ and {\it cgs} units. Here we see, that
depending on the exact parameters, the signal from the sheath moving
with $\Gamma\sim 3$ and of luminosity
$L\sim10^{48}$~erg$\cdot$sec$^{-1}$ could be delayed by a few to
few tens of seconds with respect to the GW signal. 

\section{Conclusions}

In this work, we investigate a different electromagnetic counterpart of gravitational wave events, which is the 
``off-axis" prompt emission form the associated short GRB. We argue that in a realistically structured jet, the prompt 
emission can still be detected for substantially misaligned observers and we have performed simulations to 
support this claim. Even though this prompt signal is much fainter compared to an on-axis short GRB, we stress that 
the temporal coincidence with a LIGO trigger will be crucial in order to make it a significant detection.
The localization of the ``off-axis" prompt emission $\gamma$-rays will greatly facilitate the source localization, 
host galaxy identification and detection of longer wavelength signals expected days to years after the burst. 
\section*{Acknowledgments}

We thank Alexander Tchekhovskoy for useful discussions and the anonymous referee for valuable comments on the draft. We acknowledge support from NASA through the grants NNX16AB32G and NNX17AG21G issued through the Astrophysics Theory Program. 

\bibliographystyle{mn2e}
\bibliography{references} 

\end{document}